\documentclass[prl,aps,twocolumn,preprintnumbers,showpacs,superscriptaddress]{revtex4}
\usepackage{graphicx}

\newcommand{\be}{\begin{equation}}
\newcommand{\ee}{\end{equation}}
\newcommand{\bea}{\begin{eqnarray}}
\newcommand{\eea}{\end{eqnarray}}

\newcommand{\ket}[1]{ |  #1  \rangle}
\newcommand{\bra}[1]{ \langle #1   |}
\newcommand{\proj}[1]{\ket{#1}\bra{#1}}

\begin{document}


\title{
Multipartite entanglement in quantum spin chains
 }

\author{Dagmar \surname{Bru\ss}}%
\affiliation{Institut f\"ur Theoretische Physik, %
             Heinrich-Heine-Universit\"at D\"usseldorf, %
             40225 D\"usseldorf, Germany}

\author{Nilanjana \surname{Datta}}
\affiliation{Statistical Laboratory, DPMMS,
             University of Cambridge,
             Cambridge CB3 0WB, UK }

\author{Artur \surname{Ekert}}%
\affiliation{Centre for Quantum Computation, DAMTP,
             University of Cambridge,
             Cambridge CB3 0WA, UK}
\affiliation{Department of Physics,
             National University of Singapore,
             Singapore 117\,542, Singapore}

\author{Leong Chuan \surname{Kwek}}%
\affiliation{Department of Natural Sciences, NIE, %
             Nanyang Technological University, %
             1 Nanyang Walk, %
             Singapore 637\,616, Singapore}

\author{Chiara \surname{Macchiavello}}%
\affiliation{Dipartimento di Fisica ``A. Volta"%
             and INFM-Unit\'a di Pavia,%
             Via Bassi 6, 27100 Pavia, Italy}


\begin{abstract}
We study the occurrence of multipartite entanglement in spin
chains. We show that certain genuine multipartite entangled
states, namely W states, can be obtained as ground states of
simple XX type ferromagnetic spin chains in a transverse
magnetic field, for any number of sites. Moreover, multipartite
entanglement is proven to exist even at finite temperatures. A
 transition from a product state to a multipartite
entangled state occurs when decreasing the magnetic field to a
critical value. Adiabatic passage through this 
point can thus lead to the generation of multipartite entanglement.
\end{abstract}


\maketitle


Quantum entanglement is a valuable resource with potential
applications including quantum frequency standards~\cite{BIWH96},
quantum cryptography~\cite{key} and quantum
teleportation~\cite{teleport}. Recent progress in experimental
techniques allows to generate and to control \emph{multipartite}
entanglement~\cite{expt1, expt2}. This motivates studies of
quantum entanglement in more complex physical systems, for
example chains of interacting spins. Here we show 
how passage through avoided crossings in spin chains
can be used to generate
multipartite entanglement.

Quantum spin chains have been extensively studied in the context
of quantum information science, in particular their use as quantum
wires~\cite{transfer} and as simple quantum
processors~\cite{heis}. Bipartite entanglement in spin chains has
been also thoroughly
investigated~\cite{fazio,localizable,nielsen,vedral,dorner,korepin,
paola,peng}. In
contrast, very little is known about multipartite entanglement in
such systems. A notable exception is work by Wang, who performed a
numerical analysis of multipartite entanglement in the Heisenberg
model~\cite{wang}, and by  Stelmachovic and Buzek, who showed that
in the limit of infinite coupling strength the ground state of the
Ising chain is locally unitarily equivalent to the $N$-partite
generalization of the GHZ state~\cite{buzek}. More recently Santos
showed that the presence of defects in a spin chain governed by
the XXZ Hamiltonian could be exploited to generate bipartite and
tripartite  entangled states between selected spins~\cite{santos}.

Here we analyze a quantum spin chain with an arbitrary number of
sites and demonstrate the existence of multipartite entanglement
of the W--type (see Eq.~(\ref{wn}) below) in the ground state of
the XX Hamiltonian with suitable transverse magnetic field. We
find the critical value of the magnetic field at which the ground
state undergoes a  transition from a product state to
this multipartite entangled state. Our work therefore provides a
method for generating a multipartite W--type entangled state by
tuning a {\em{single}} {\em external} parameter, namely the global
magnetic field, while the inter-spin interaction remains constant.
We refer to spins as qubits and use $\ket{0}$ and $\ket{1}$ to
denote spin-up and spin-down states, respectively.

Let us  start with a simple example, where the transition from a
product state to a multipartite entangled state can be easily
seen. We consider the XXZ model for a chain of $3$ spins in an
external magnetic field and with periodic boundary conditions. The
corresponding Hamiltonian is given by
\be%
H_{{\rm \tiny XXZ}}=\sum_{i=1}^3 \left(J\vec \sigma_i \otimes
\vec \sigma_{i+1} + \Delta \sigma_i^z \otimes\sigma_{i+1}^z 
+
 B\,\sigma_i^z\right)\ , \label{xxz}
\ee%
where $J$ is the coupling in the $x$ and $y$ directions and $J +
\Delta$ is the coupling in the $z$ direction. The parameter
$\Delta$ quantifies the anisotropy in the interaction; for the
Heisenberg interaction $\Delta=0$. N.B. this model exhibits the
same symmetry as the XX model with $N$ sites which will be
discussed later. In both cases the 
$z$-component of the
total spin, $\sigma^z_{tot} =
\sum_{i} \sigma^z_i$, commutes with the  Hamiltonian $H$.
Thus, the eigenstates of $H$ are superpositions of states with a
fixed number of up-spins.

In Eq.~(\ref{xxz}) the term proportional to $\Delta$  commutes
with the terms proportional to $J$. Thus, the eigenstates of the
Hamiltonian $H_{XXZ}$ coincide with those of the isotropic
Heisenberg model. There are four non-degenerate eigenstates, given
by $\ket{111},\ket{\bar
W}=(\ket{110}+\ket{101}+\ket{011})/\sqrt{3},
\ket{W}=(\ket{001}+\ket{010}+\ket{100})/\sqrt{3}$, and
$\ket{000}$. The corresponding eigenvalues are
$E_{{111}}=3J+3B+3\Delta$, $E_{{\bar W}}=3J+B-\Delta$,
$E_{{W}}=3J-B-\Delta$, and $E_{{000}}=3J-3B+3\Delta$. The
remaining eigenstates are doubly degenerate and are given by
$\ket{\bar W^{(k)}}=(\ket{110}+e^{2\pi i k/3}\ket{101}+e^{-2\pi i
k/3}\ket{011}) /\sqrt{3}$ with $k=1,-1$ and eigenvalue $E_{{\bar
W^{(k)}}}=J+B-\Delta$, and $\ket{W^{(k)}}=(\ket{001}+e^{2\pi i
k/3}\ket{010}+e^{-2\pi i k/3}\ket{100}) /\sqrt{3}$ with eigenvalue
$E_{{W^{(k)}}}=J-B-\Delta$.

The relative values of the parameters $J,B$ and $\Delta$ determine
which of these states is the ground state. Let us first consider
the isotropic case, $\Delta=0$. When the value $J$ is negative,
i.e. in the ferromagnetic case, the ground state is always a
product state, regardless of the value of $B$. Changing the sign of
the magnetic field $B$ simply leads to the relabeling
$0\leftrightarrow 1$ of each qubit. 
In the following, we will always consider $B$ to be positive.
For positive $J$, i.e. the
antiferromagnetic case, and for $B>J$ the ground state is also a
product state. However, for $B<J$ the ground state lies in the
subspace spanned by $\ket{W_k}$, with $k=1,-1$. This subspace
contains states with W-type tripartite entanglement and
biseparable states, i.e. states in which one qubit is unentangled.

\begin{figure}
\begin{center}
\hspace*{-2.5cm}
\includegraphics[width=9cm]{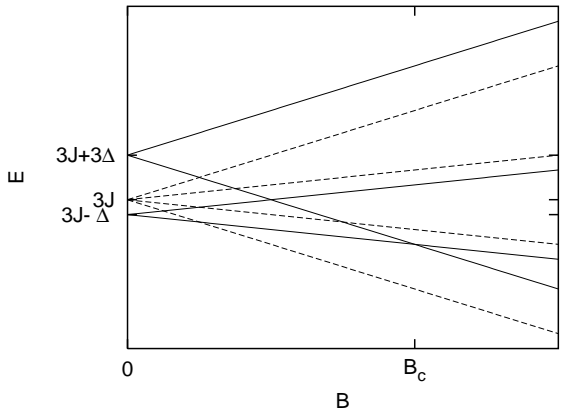}
\caption{Role of the anisotropy: Energy levels for the
ferromagnetic XXZ model with anisotropy $\Delta$ as a function of
the magnetic field $B$, for the case $N=3$. Dashed lines are for
$\Delta=0$ and solid lines for $\Delta>0$. For $B>B_c$ the levels
are ordered from bottom to top as $E_{{000}}, E_{{W}}, E_{{\bar
W}}, E_{{111}}$ (see main text for the definition), both for
$\Delta=0$ and $\Delta>0$.}
\label{levels}
\end{center}
\end{figure}

Let us now turn to the anisotropic case, $\Delta\neq 0$. For
negative values of $J$ and negative $\Delta$, the ground state is
again given by a product state, regardless of the value of $B$.
However, the case of negative $J$ and positive $\Delta$ leads to
interesting results. For high values of $B$, the spins are
aligned, and the ground state is $\ket{000}$, i.e. a product
state. When one decreases the value of $B$, at 
$B_c=2\Delta$ the ground state changes to $\ket{W}$, namely a
genuinely tripartite entangled state. The essential role of the
non-vanishing anisotropy $\Delta$ for the existence of the ground
state transition from a product state to a tripartite entangled
one is illustrated in Fig. \ref{levels}.

The above considerations hold if the spin chain is at zero
temperature. However, at finite but low temperatures, the state of
this three--spin chain can be shown to be still genuinely
tripartite entangled. To see this, one expands the density matrix
of the system at a temperature $T$, which is given by its Gibbs
state, as follows
\begin{eqnarray}
\rho &\propto & e^{-\beta E_{000}} |000\rangle \langle 000| +
e^{-\beta E_{W}} |W\rangle \langle W| \nonumber\\
&+& e^{-\beta E_{\bar W}} |\bar W\rangle \langle \bar W|+
e^{-\beta E_{111}} |111\rangle \langle 111|+ \ldots \label{exp}\ ,
\end{eqnarray}
where $\beta = 1/kT$ and $k$ is the Boltzmann constant. In the
vicinity of $B_c=2\Delta$, the two lowest eigenvalues are
$E_{000}$ and $E_{W}$. Hence by retaining only the two leading
terms of the expansion (\ref{exp}), 
 which is a good approximation for sufficiently low temperatures,
the state of the system can be
approximated 
 as
\be%
\varrho = p\proj{W}+(1-p)\proj{000}\ , \label{bmw}
\ee%
where $p=1/(1 + \exp[\mp 2\beta\,(2\Delta-B)])$
for $B_c<2\Delta$ and $B_c>2\Delta$,
respectively. This state was first studied
 in the context of approximate quantum
cloning~\cite{BM}. It is a genuinely tripartite entangled mixture,
which cannot be written as a mixture of biseparable states for any
value of $p\neq 0$. To prove this, let us assume the contrary,
namely that $\varrho$ is biseparable. A biseparable state
$\varrho$ of a tripartite system $ABC$ can be decomposed as
\cite{bisep} \be \varrho=\sum_{ijk} \left(p_i^C\tau^{AB}_i\eta^C_i
+ p_j^B\tau^{AC}_j\eta^B_j +p_k^A\tau^{BC}_k \eta^A_k \right),
\label{prf} \ee where $p_i^{C}$, $p_j^{B}$, $p_k^{A}$ are
probabilities with $\sum_{ijk}(p_i^C + p_j^B + p_k^A)=1$. In the
first term on the RHS of (\ref{prf}), $\tau^{AB}_i$ denotes the
joint density operator of the subsystems A and B, while $\eta^C_i$
denotes the density operator of the subsystem C (and analogously
for the two subsequent terms). The decomposition (\ref{prf})
implies that there exists a biseparable state of the form
$\ket{\psi^{AB}_i} \ket{\phi^C_i}$ in the range of $\varrho$. This
in turn implies that there should exist non--zero coefficients
$\alpha$ and $\beta$ such that $\ket{\psi^{AB}_i}
\ket{\phi^C_i}=\alpha\ket{W} +\beta\ket{000}$. However, it is
straightforward to see that there are no biseparable vectors in
the subspace spanned by $\ket{W}$ and $\ket{000}$, apart from
the vector  $\ket{000}$, which however corresponds to the 
trivial case $\alpha =0$. Hence one
arrives at a contradiction. Let us already note here that this
argument can be generalised in  a straightforward way to $N$
parties: any mixture of an N-party $W$ state and the product state
$\ket{00...0}$ is genuinely multipartite entangled.

Let us now turn to  the general case of a spin chain with $N$
sites. Here we will focus on the choice $\Delta = -J$, namely the
XX model,
with periodic boundary conditions, described by the Hamiltonian
\be%
H_{{\rm\tiny XX}} =\sum_{i=1}^{N}( J (\sigma_i^x
\otimes\sigma_{i+1}^x + \sigma_i^y \otimes\sigma_{i+1}^y) +
B\,\sigma_i^z ) \ . \label{hxy}
\ee%
Again, we consider the ferromagnetic case $J<0$, and $B>0$. As in
the case of the XXZ model with three sites, $[H_{XX},
\sigma^z_{tot}] = 0$ holds, where $\sigma^z_{tot} = \sum_{i=1}^N
\sigma^z_i$. Hence, the Hilbert space decomposes into invariant
subspaces, each corresponding to a distinct eigenvalue of
$\sigma^z_{tot}$ which we denote as $m$. It represents the total
number of 
excitations (down-spins) in the chain. There are $N+1$ such subspaces,
corresponding to spin configurations with $m= 0,1,2,\ldots,N$. We
refer to the subspace corresponding to a particular $m$ as the
$m$--excitation subspace. The subspaces for which $m= 0$ and $m=N$
are one-dimensional. The eigenstates of the Hamiltonian $H$ in
these subspaces are given by the product states $|000\ldots
0\rangle$ and $|111\ldots 1\rangle$, respectively. The
corresponding eigenvalues are
\be%
 E^{(0)} = - NB \hspace{1cm}
\label{en1} \mbox{and} \hspace{1cm}
 E^{(N)} = + NB \ .
\ee%
The energy eigenvalues for the $m$-excitation subspace can be
obtained using the standard fermionization technique, introduced
by E. Lieb {\em et al} and S. Katsura~\cite{ferm}. 
The eigenvalues are labeled by $m$
distinct quantum numbers $k_1, k_2, \ldots, k_m \in \{1,2,,\ldots,
N\}$ and can be expressed as
\bea%
E^{(m)}_{ k_1, k_2,\ldots, k_m}& =& 4J\Bigl[ \cos \bigl(\frac{2
\pi k_1}{N}\bigr)
+ \cos \bigl(\frac{2 \pi k_2}{N}\bigr) + \ldots \nonumber \\
&&+ \cos \bigl(\frac{2 \pi k_m}{N}\bigr)\Bigr] - (N-2m) B.
\label{en5}
\eea%
In particular, for the single excitation subspace,
\be%
E^{(1)}_k = 4J \cos (\frac{2\pi k}{N}) - (N-2)B, \label{en2}
\ee%
where $k=1,\ldots, N$. The corresponding energy eigenstates are
given by
\be%
|\phi_k\rangle = \frac{1}{\sqrt{N}} \sum_{n=1}^N
e^{2\pi i k n/N} |n\rangle \label{state}
\ee%
where the states $|n\rangle$, with $n=1,2,\ldots,N$, correspond to
spin configurations in which all spins are up, apart from the
spin at the site $n$ which is down. For example, for $N=3$ we have
$\ket{1}=\ket{100}$, $\ket{2}=\ket{010}$ and $\ket{3}=\ket{001}$.
These states form a complete basis in the single excitation
subspace.

Let us now look at the ground state of the chain as a function of
the magnetic field $B$. If the chain is held in a very strong
magnetic field $B\gg J$ then its ground state is the product state
of the form $|000\ldots 0\rangle$ and energy $E^{(0)}=- NB$.
However, when we decrease $B$, keeping $J$ fixed, then the lowest
energy single excitation eigenstate $\ket{\phi_N}$, which from now
on we will label as $\ket{W_N}$,
\be%
|W_N\rangle = \frac{1}{\sqrt{N}} \Bigl[ |1\rangle + |2\rangle +
|3\rangle + \ldots + |N\rangle \Bigr], \label{wn}
\ee%
becomes the new ground state with energy $E^{(1)}_N=4J - (N-2)B$.
The crossover 
 from the product state
$|000\ldots 0\rangle$ to the entangled state $\ket{W_N}$ occurs at
the critical value of the field $B_c=-2J$, which is independent of
$N$. The new ground state $|W_N\rangle$ is a generalization of the
tripartite entangled $W$--state to $N$ spins and is genuinely
$N$--partite entangled.

This is the first crossover because in our particular case of
$J<0$ the lowest energy in the $m$-excitation subspace (for $m\ge
2$) is lower bounded by $4Jm - (N-2m)B$ (see Eq.~(\ref{en5})),
which can be written as $E^{(1)}_N + 2(m-1)(2J +B)$ and is
certainly greater than $E^{(1)}_N$ for all values of $B \ge -2J$ .

Note that the value $B_c=-2J$ corresponds to the 
critical point for the  Bose Hubbard model
with infinite on-site repulsion energy of the
bosons~\cite{sachdev}, namely to the quantum phase transition from the 
Mott
insulator to superfluid phase. We also point out that, as explained below
Eq.~(\ref{bmw}), at finite temperatures and near $B_c=-2J$
 there remains genuine multipartite entanglement.

In order to generate the state $|W_N\rangle$ from the state
$|00\ldots 0\rangle$ by lowering the magnetic field $B$ from an
initial large value, it is necessary to turn the level crossing at
$B_c=-2J$ into an avoided crossing. An avoided
crossing can be realized by adding a small  perturbation to
the Hamiltonian $H_{XX}$, e.g., a term ${B'\sum_{i=1}^N
\sigma_i^x}$, with $B' << B$. In the presence of an avoided
crossing, a transition from $|00\ldots 0\rangle$ to $|W_N\rangle$
can be achieved by lowering the magnetic field slowly enough, so
that the ground state adiabatically follows~\cite{adia}. The
required rate of change of $B$ depends inversely on the gap of the
avoided crossing. Using degenerate perturbation theory, this gap
is found in first order to be equal to $2B'\sqrt{N}$.

Before concluding, let us make a brief remark about the ground
state for values of the magnetic field in the range $B<-2J$. It
turns out that the ground state changes successively from one
excitation number to the next-higher one when decreasing the value
of the magnetic field $B$. To show this, let us calculate  the
crossings of the energy levels for different excitation numbers
(for large  $N$). It is straightforward to see that the lowest
energy in the $m$--excitation subspace, for {\em{odd}} values of
$m$, is given by
\be%
\widetilde{{E}_{\rm{odd}}^{(m)}} = 4J\Bigl[ 1
+ \sum_{j=1}^{[m/2]} 2 \cos \bigl(\frac{2 \pi j}{N}\bigr)\Bigr] -
(N-2m) B \, \label{11}
\ee%
whereas, for {\em{even}} values of $m$, it is given by
\be%
\widetilde{{E}_{\rm{ev}}^{(m)}} = 4J\Bigl[ 1 + \sum_{j=1}^{m/2-1} 2 \cos
\bigl(\frac{2 \pi j}{N}\bigr) + \cos\bigl(\frac{\pi m}{N}\bigr)\Bigr]
- (N-2m) B.
\label{22}
\ee%
From this it follows that the crossing  $\widetilde{{E}^{(m+1)}}
=\widetilde{{E}^{(m)}}$ occurs at $B_c^{(m)}=-2J[1-(2m^2
\pi^2)/N^2]$. Thus, with the addition of a suitable  small
perturbation, and by adiabatically decreasing the magnetic field,
transitions through a cascade of ground states with increasing
excitation numbers can be obtained.

In conclusion,  we have shown that a multipartite entangled state
of the W-type occurs naturally as a ground state in the
ferromagnetic XX spin chain with $N$ sites in an external magnetic
field. 
W-states are a useful
resource for quantum information processing tasks, e.g.
quantum teleportation and dense coding \cite{teleportw}.
Our analysis suggests a new method of generating an
$N$-partite entangled W-state, namely by driving the chain
adiabatically through 
 an avoided level crossing.
This amounts to preparation of the initial product
state $|000\ldots 0\rangle$ in a strong magnetic field, $B\gg J$
and then slowly reducing the strength of the field, in the presence
of a small 
perturbation, until it
passes through the  value $B_c=-2J$ and generates the
$N$-partite entangled W-state. (Let us mention in passing that the
concurrence, which measures entanglement between any two qubits in
the chain, has the value $2/N$. In this sense the W state also carries
bipartite long-range entanglement.) This method does not require
any dynamical control over the couplings in the spin chain. Only
one external global parameter, namely the magnetic field $B$, has
to be modified. Thus, our result opens new possibilities for the
creation of multipartite entanglement in condensed matter physics.
Moreover, if each qubit in the chain can be controlled separately
then this transition can be achieved by local operations, i.e. by
reducing the field locally at each on the $N$ sites of the chain.
This does not imply that entanglement can be created by local
operations as qubits do interact with each other, but it opens new
possibilities of manipulating multipartite quantum entanglement.

We wish to thank M. Lewenstein and  M. Palma for discussions. This
work was supported in part by a grant from the Deutsche
Forschungsgemeinschaft, the Cambridge-MIT Institute, A$^*$Star
Grant No.\ 012-104-0040 and the EU under project RESQ and
QUPRODIS. We acknowledge support from the Benasque Center of
Science during the Workshop on Quantum Information and
Communication, 2003.

\end{document}